\documentclass{osa-article}

\journal{osac}

\articletype{Research Article} 
\usepackage{amssymb}
\usepackage{graphicx}
\usepackage{caption}
\usepackage{subcaption}
\usepackage{multirow}
\usepackage{soul}
\usepackage{kantlipsum}
\usepackage{amsmath}
\usepackage{amsfonts}
\usepackage{amssymb}
\usepackage{dcolumn} 
\usepackage{braket}
\usepackage{booktabs}
\usepackage{multirow}
\usepackage[table]{xcolor}
\usepackage{dsfont}
\usepackage{times}
\usepackage{epsfig}
\usepackage{xcolor}
\usepackage{tikz}
\usepackage{float}
\usepackage{color}
\usepackage{inputenc}
\usepackage{booktabs}
\usepackage{pgfplotstable}
\def\bx {{\bf x}}

\def\t  {{\theta }}

\def\bkappa {\boldsymbol{\kappa}}

\def\o {{\omega}}

\newcommand{\beq}{\begin{equation}}
\newcommand{\eeq}{\end{equation}}
\newcommand{\ba}{\begin{eqnarray}}
\newcommand{\ea}{\end{eqnarray}}

\newcommand{\kp}{{\bf k}\mkern -3mu\cdot \mkern -3mu{\bf p}}
\def\cE {{\mathcal E}}

\def\cH {{\mathcal H}}
\def\bkn{{n\bkappa}}

\def\bkmprime{{m\bkappa^\prime}}

\input epsf

\usepackage[english]{babel}

\begin{document}

\title{Novel topological beam-splitting in photonic crystals}

\author{Mehul Makwana,\authormark{1,2*} Richard Craster,\authormark{1} and S\'ebastien Guenneau\authormark{3}}

\address{\authormark{1}Department of Mathematics, Imperial College London, London SW7 2AZ, UK\\
\authormark{2}Multiwave Technologies AG, 3 Chemin du Pr\^e Fleuri, 1228, Geneva, Switzerland\\
\authormark{3}Aix$-$Marseille Univ, CNRS, Centrale Marseille, Institut Fresnel, Marseille, France}

\email{\authormark{*}mehul.makwana07@imperial.ac.uk} 



\begin{abstract}
We create a passive wave splitter, created purely by geometry, to engineer three-way beam splitting in electromagnetism in transverse electric and magnetic polarisation. We do so by considering arrangements of Indium Phosphide dielectric pillars in air, in particular we place several inclusions within a cell that is then extended periodically upon a square lattice. Hexagonal lattice structures are more commonly used in topological  valleytronics but, as we discuss, three-way splitting is only possible using a square, or rectangular, lattice. To achieve splitting and transport around a sharp bend we use accidental, and not symmetry-induced, Dirac cones. Within each cell pillars are either arranged around a triangle or square; we demonstrate the mechanism of splitting and why it does not occur for one of the cases. The theory is developed and full scattering simulations demonstrate the effectiveness of the proposed designs. 
\end{abstract}

\section{Introduction}

Beam-splitters play a key role in many optical devices, in particular interferometers, with a consequently broad variety of applications ranging from astrophysics \cite{quirrenbach01a} to quantum computing \cite{kok07a} as well as to many areas of optical electronics such as optical modulators \cite{mitomi95a} for fibre optic telecommunications amongst much more. Indeed to achieve complete control over the flow of light, power division and redirection devices are required, of which beam-splitters are those most commonly utilised; a recurrent theme is the desire to have broadband lossless splitters, capable of multiple and tunable re-direction, and with dimensions that are sub-micron, or at most microns, in size. 

Several different beam-splitting approaches have been successfully implemented ranging from coupled bent dielectric slab waveguides or ridge waveguides atop  substrates, to photonic crystal/ grating devices using waveguide or self-collimation or more recently non-reciprocal media or using topological designs. The dielectric waveguide approach is exemplified for polarisation beam splitters, as used for Mach-Zehnder interferometers, by \cite{soldano94a,mitomi95a} and recently with more compact designs \cite{dai11b}. Splitters based upon dielectric waveguides traditionally  
have disadvantage of scale, even for high-index dielectrics requiring them to be of the order of several wavelengths in length to minimise scattering and losses at corners or shallow bends. 
This scale limitation, and a desire to minimise radiation losses at the bend, motivated \cite{mekis96a} to take advantage of photonic waveguides created by removing rows or columns of the crystal array; here we design the topological analogues. These photonic waveguide devices have been successfully applied to  beam-splitters and branched waveguides with  
T-shaped and Y-shaped branches implemented by \cite{bayinder00a,fan01a,boscolo02a}, amongst many others, and the insertion of additional, or modified, array elements in the neighbourhood of the corner \cite{chutinan02a} can improve transmission or broaden frequency range; this has spawned ever more elaborate designs to optimise for various scenarios \cite{jensen04a,erol15a}. A related, but different, approach to creating splitters within photonic crystals is to take advantage of self-collimation \cite{yu03a} and self-guiding created by strong dynamic anisotropy 
 \cite{Chigrin:03}. Beam splitters, including three-way splitters \cite{pustai04a}, are created and applied to devices 
 \cite{prather07a,zhao07a} using lattice array alterations to create effective  mirrors or partial mirrors. 

 
More recently, ideas from topological insulators \cite{kane_z2_2005} have been transposed into photonics \cite{lu_topological_2014}, as reviewed in  \cite{khanikaev_two-dimensional_2017}, showing promise for robust one-way edge states protected against disorder by topology. This promise is tempered by the requirement for time reversal symmetry (TRS) to be broken, and the electron spin to be mimicked  by pseudospin in continuum systems. A simpler, passive and time reversal symmetric,  but less robust approach is to attempt to reproduce the valley-Hall effect and utilise ideas from the field of valleytronics \cite{xiao_valley-contrasting_2007-1}, for instance \cite{ma_all-si_2016} create dielectric photonic topological arrangements leading to reflectionless guiding and designs for optical delay lines. Topological designs have been recently implemented for telecommunication wavelengths \cite{shalaev18a} on a CMOS-compatible chip thus bringing these concepts closer to application. These valley-Hall devices are locally topologically nontrivial however globally trivial, and therefore cannot draw upon the full power of the analogy with topological insulators, but do have advantages in terms of simplicity of construction as one need only break spatial inversion or reflectional symmetry, together with suppressing backscatter. Given the emergence of topological guiding there is now interest in developing this for photonic circuits, \cite{zhang18a}, and a natural drive to explore the potential of these new ideas. The vast majority of this 
valleytronics literature takes advantage of periodic hexagonal or honeycomb lattices, utilising ideas from graphene, in particular the symmetry properties of the hexagonal Brillouin zone and symmetry induced Dirac cones at the $KK^\prime$ vertices. Perturbations of the structure, that break symmetries, then gap the Dirac points giving  topologically nontrivial band-gaps and well-defined $KK'$ valleys. The valleys have opposite chirality and are related by parity and/or reflectional symmetry as well as TRS. A key point is that by engineering a large Fourier separation between the valleys one suppresses intervalley scattering and ultimately the valley is used as an information carrier  \cite{khanikaev_two-dimensional_2017}. One negative that emerges is geometrical: 
 The hexagonal systems can only create two-way energy-splitters \cite{makwana18b}, and three-way splitting would require other geometries for which symmetry induced Dirac points do not occur--hence the recipe outlined above cannot be employed. 
 
\begin{figure}[ht!]
\mbox{}
\includegraphics[width=5.5cm]{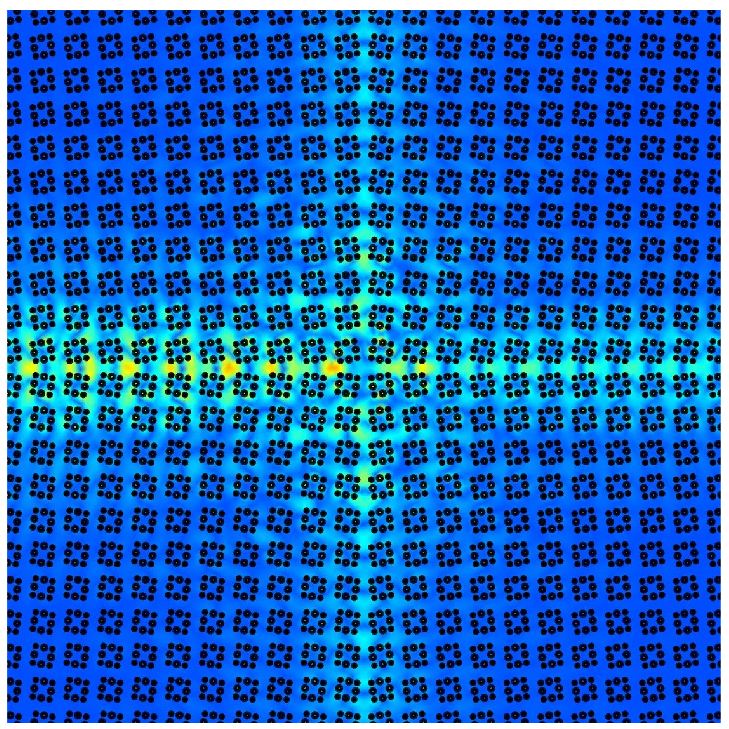}
\hspace{1cm}
\includegraphics[width=5.5cm]{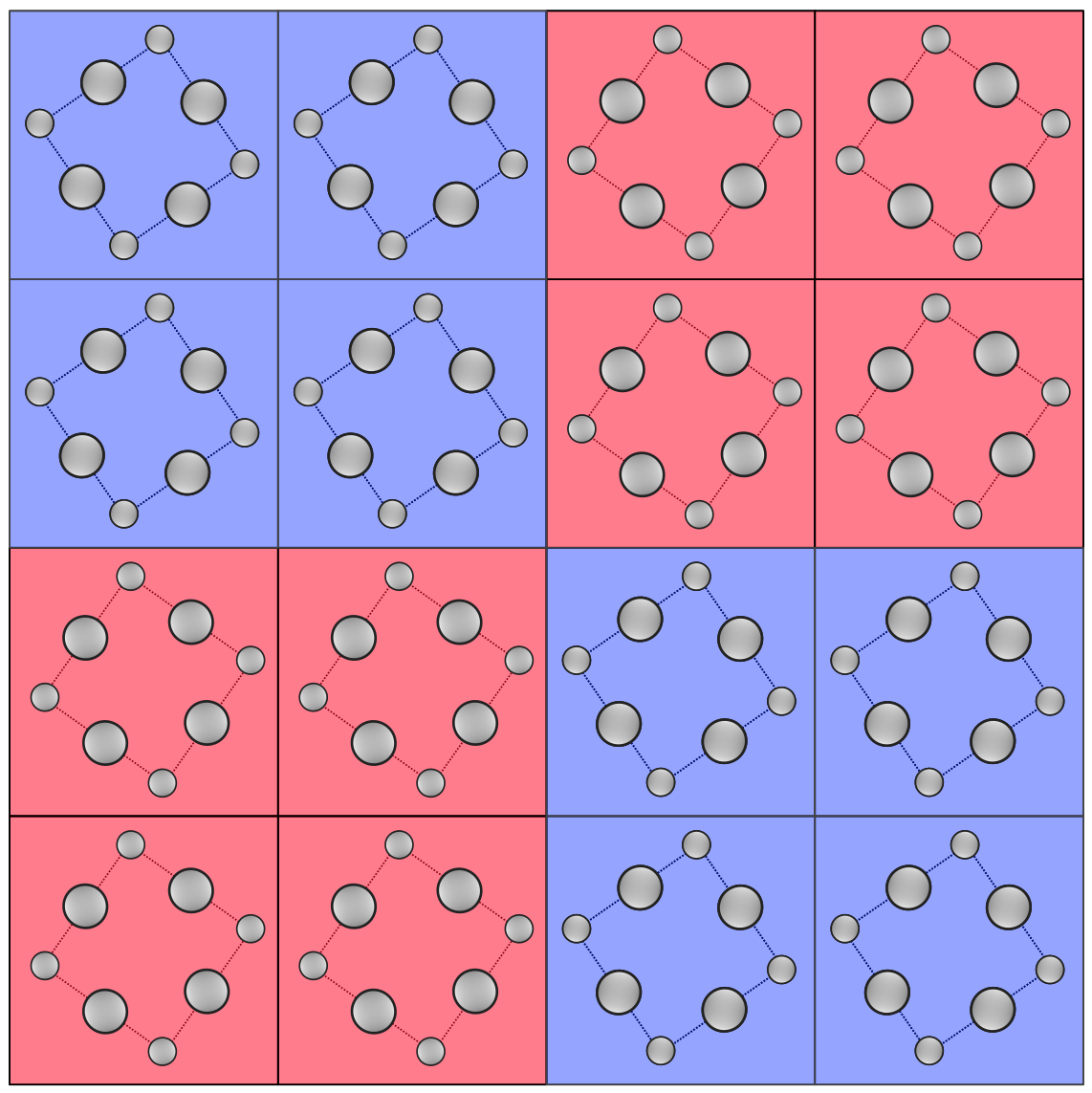}
\caption{A three-way splitter designed on a square lattice using accidental degeneracies and geometry (right panel) shows the junction region between four quadrants each containing a set of dielectric inclusions placed around a rotated square. The quadrants differ in only the rotation given to the inclusion arrangement within it. In the left panel, the wave energy incoming along the leftmost interface is split at the junction into three interface waves. In this simulation the normalized angular frequency $\omega d/c$ is $4.78$ and we treat TM polarisation. 
}
\label{fig:motivation}
\end{figure}

Fortunately, one can engineer accidental Dirac points \cite{sakoda14a,he15a} for square systems (and indeed for all two-dimensional systems, most usefully for those that do not have symmetry induced Dirac cones) and, although they are no longer at the high symmetry points one can mimic some of the effects found in hexagonal systems for the usual symmetry induced Dirac points, see for instance \cite{xu16a}. Recently \cite{makwana19a} showed beam splitting for a model system of waves on elastic plates, for a fourth order elastic plate equation and masses of infinitesimal radius, here we explore to what extent three-way beam splitters can be employed in electromagnetism. We do this via a combination of group and $\kp$ theory, \cite{sakoda12a}, coupled with detailed numerical simulations to extract the interfacial edge states, zero-line modes (ZLMs), and further we take these idealised edge states and perform numerical simulations,  using \cite{comsol}, showing transport around right-angled bends and three-way energy splitting. 
A typical splitter that we construct is shown in Fig. \ref{fig:motivation}, it is constructed from dielectric inclusions arranged upon a square lattice, within each elementary cell there is an arrangement of inclusions and the choice of arrangement is critical for splitting; we illustrate this importance by contrasting two arrangements. A judicious choice of inclusion arrangements, and the connection of quadrants of material yield the passive splitter that, as in Fig. \ref{fig:motivation}, takes an incoming wave and then splits it in three. 

\begin{figure}[!htb]
    \centering
    \captionsetup{justification=raggedright}
        \begin{minipage}{.25\textwidth}
         \caption*{(a) $C_{3v}$ }
        \includegraphics[width=0.85\linewidth]{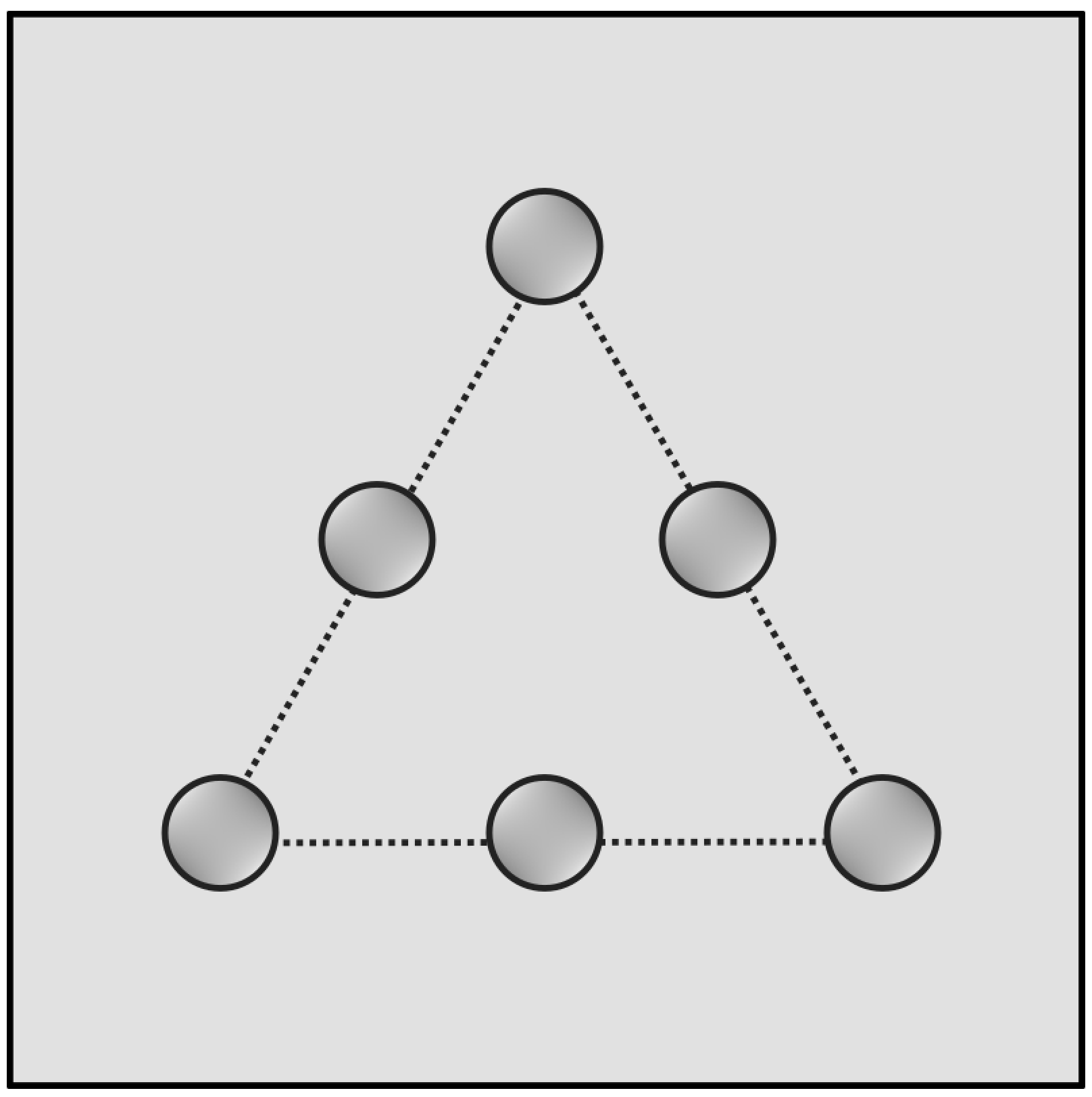}
        \label{fig:prob1_6_1}
    \end{minipage}%
    \begin{minipage}{0.25\textwidth}
         \caption*{(b) $C_{4v}$}
        \includegraphics[width=0.85\linewidth]{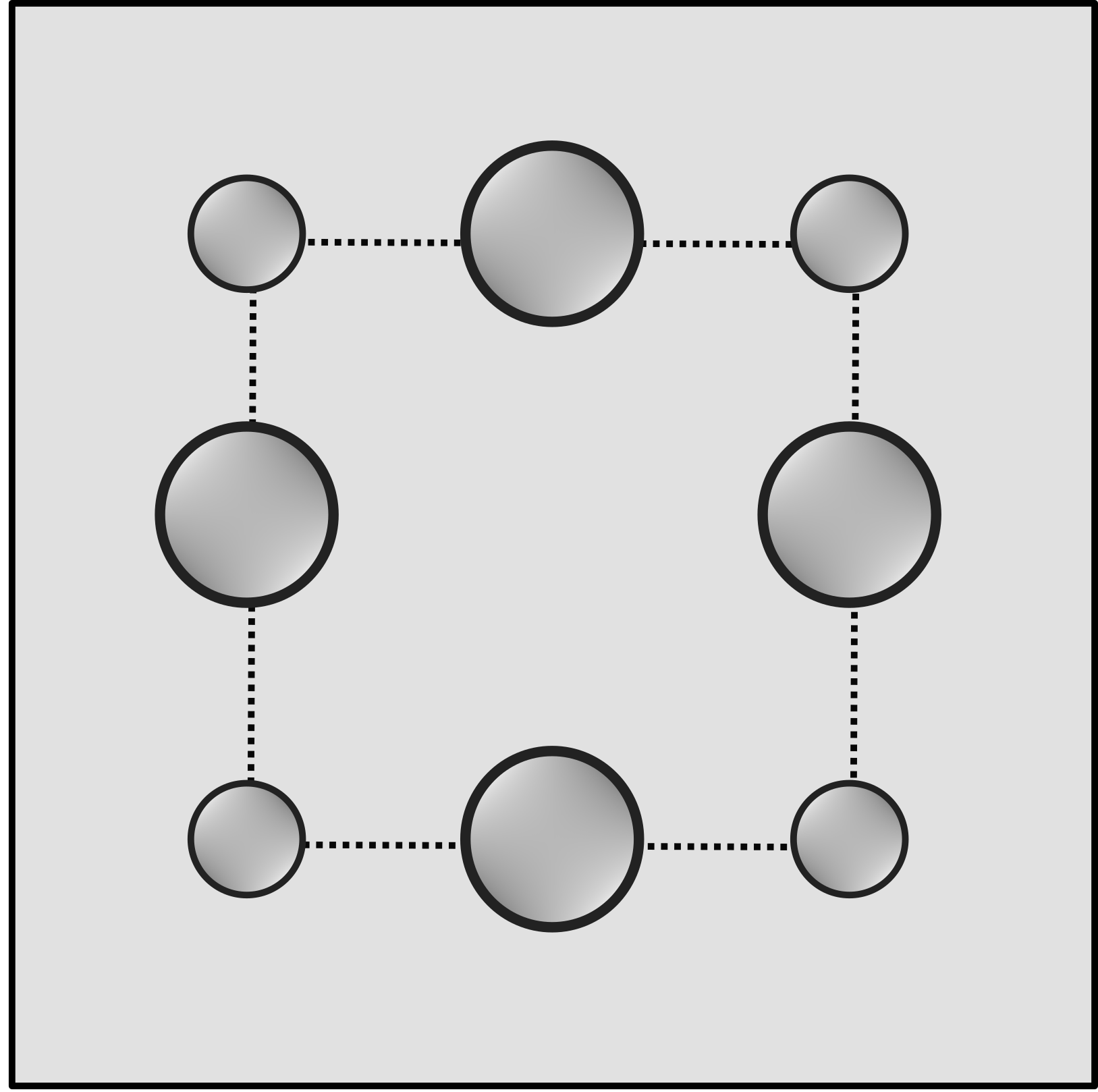}
        \label{fig:prob1_6_2}
    \end{minipage}
           \begin{minipage}{0.25\textwidth}
         \caption*{\vspace{-0.25cm}(c)}
        \includegraphics[width=0.95\linewidth]{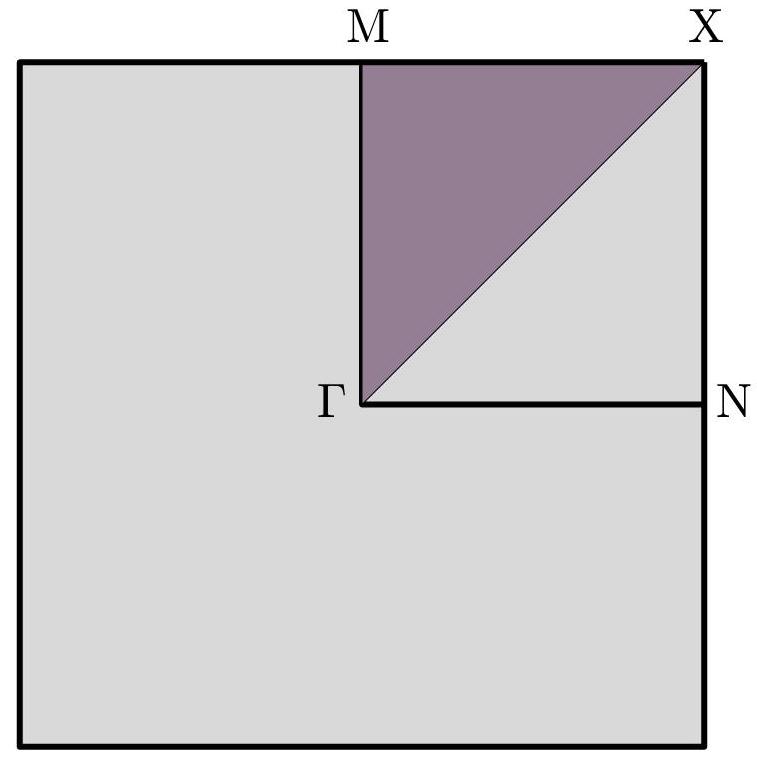}
        \label{fig:prob1_6_3}
    \end{minipage}
    \caption{The (non-rotated) arrangement of inclusions taken in the elementary cell: (a) equilateral triangular, $C_{3v}$, (b) square, $C_{4v}$, arrangements and (c) the first Brillouin zone relevant for periodic arrangements of these cells upon a square lattice. For a pitch $d$, the radius of inclusions is $0.05d$ and their center-to-center spacing is $0.2$ in the triangular case (a) 
the radii of the two types of inclusions in the square case are $0.075d$ and $0.05d$ and their center-to-center spacing is $0.25d$ The high-symmetry points of the Brillouin zone we refer to are $\Gamma=(0,0)$, $N=(\pi/d,0)$, $M=(0,\pi/d)$ and $X=(\pi/d,\pi/d)$.}
    \label{fig:Cells_IBZ}
\end{figure}

We operate at  telecommunication wavelengths using a photonic crystal (PC) consisting of Indium Phosphide (InP) dielectric pillars in air. Drawing upon the design and fabrication of a dielectric carpet in \cite{Scherrer:13}, we contrast two designs of PCs within a square lattice array; the first of which has 6 InP dielectric pillars of diameter $200$ nm, height $2\mu$m, and minimum center-to-center spacing of $250$nm arranged around an equilateral triangle, Fig. \ref{fig:Cells_IBZ}(a), and the second has 8 pillars with similar dimensions arranged around a square as in Fig. \ref{fig:Cells_IBZ}(b). The refractive index of InP is $n=3.16$, and we consider primarily (TM) polarization whereby the electric field is parallel to the pillars' axis and discuss the analogous transverse electric (TE) polarization, the magnetic field is parallel to the pillar axis, in the concluding remarks. These two designs exemplify the potential of topological designs for energy splitting, and in particular how to achieve three-way splitting, for intrachip communication devices.

The Maxwell equations split naturally into $p$ and $s$ polarizations, with forcing created by an electric line source or magnetic current dipole at position ${\bf r_s}$ respectively, \cite{zolla07a}, as TM  
\beq 
\nabla \times (\mu_r^{-1}\nabla\times {\bf E}_l)-\varepsilon_r\mu_0\epsilon_0\omega^2{\bf E}_l=-i\omega I_s \mu_0\delta_{\bf r_s}{\bf e}_3,
\eeq 
and TE,
\beq 
\nabla \times (\epsilon_r^{-1}\nabla\times {\bf H}_l)-\mu_r\mu_0\epsilon_0\omega^2{\bf H}_l=\nabla\times (\epsilon_r^{-1}{\bf j}_T)
\eeq 
with $\epsilon_0,\mu_0$ ($\epsilon_r,\mu_r$) as the permittivity and permeability in-vacuo (and relative values), $\omega$ is frequency with time-harmonic waves, $\exp(-i\omega t)$, assumed, ${\bf j}_T$ and $I_s$ being currents. We will often use a normalised frequency, $\omega d/c$, hereafter where $d$ is the pitch of the lattice and $c$ the speed of light, $c^2=1/\mu_0\epsilon_0$. 
The TM field is driven by a line monopole source at $\delta_{\bf r_s}$ and the TE field by a line dipole. For the polarised fields, taking a Cartesian coordinate system $(x_1,x_2,x_3)=({\bf x},x_3)$, where ${\bf x}$ is the in-plane (or transverse) variable and $x_3$ is the out-of-plane (or longitudinal) variable, we use invariance  along $x_3$ to split the vector Maxwell system into two polarizations: The TM polarization has ${\bf{E}}_l=\cE({\bf x}){\bf e}_3$ (TE similarly has ${\bf{H}}_l=\cH({\bf x}){\bf e}_3$), that is, the field is perpendicular to the pillars, and we concentrate from hereon on the TM polarized field. We take the relative permittivity to be spatially dependent, i.e. $\epsilon_r\equiv \epsilon_r(\bx)$, with the remaining parameters constant, and thus for source-free TM fields we have that  
\beq
\nabla^2
\cE({\bf x}) + a({\bf x})\omega^2\cE({\bf x})=0
\label{eq:TM}
\eeq
where $\nabla^2$ is Laplacian with respect to the in-plane variable ${\bf x}$, and the material dependence is encapsulated in $a(\bx)=\mu_0\epsilon_0\mu_r\epsilon_r(\bx)$. 
Eq. \eqref{eq:TM} carries the assumption of material isotropy and scalar permittivity and permeability that is lossless. Moreover, for the pillar crystal $a(\bx)$ is a piecewise constant function as $\varepsilon_r$ is $1$ in air and $3.16^2$ in the pillars, and $\mu_r=1$ everywhere (a non-magnetic medium). This is equivalent to Helmholtz equations in each homogeneous phase, coupled through continuity conditions across the pillar interfaces of the field $\cE$ and its normal derivative $\partial n \cE= {\bf n}\cdot\nabla \cE$, with ${\bf n}$ the normal to the interface.

\section{Engineering a non-symmetry repelled Dirac cone for a square lattice}
To create three-way splitting we first need to engineer accidental degeneracies along the Brillouin zone as shown in Fig. \ref{fig:Cells_IBZ}(c), boundary, it is not sufficient to simply generate Dirac points, and the arrangement of inclusions in the interior of the physical elementary cell, and their symmetry, plays a key role. We utilise perturbation theory and symmetry arguments to design the structures we study; the inclusions are either in a triangular formation (and have $C_{3v}$ symmetry) or are arranged along a square (and have $C_{4v}$ symmetry), see Fig. \ref{fig:Cells_IBZ}(a,b). As we shall see the interaction of the internal inclusion symmetries with the symmetries of the lattice lead to fundamentally different behaviours; ultimately we shall see that the $C_{3v}$ case is unable to create splitters for reasons uncovered by symmetry arguments. We begin by determining the criteria under which accidental Dirac points are created, then gap them using geometrical rotations that break symmetry to create band-gaps with protected edge states; these edge states form the building blocks of the splitters we design. 

\subsection{Perturbation theory and band interactions}

\begin{figure}[ht!]
\mbox{}
\includegraphics[width=12cm]{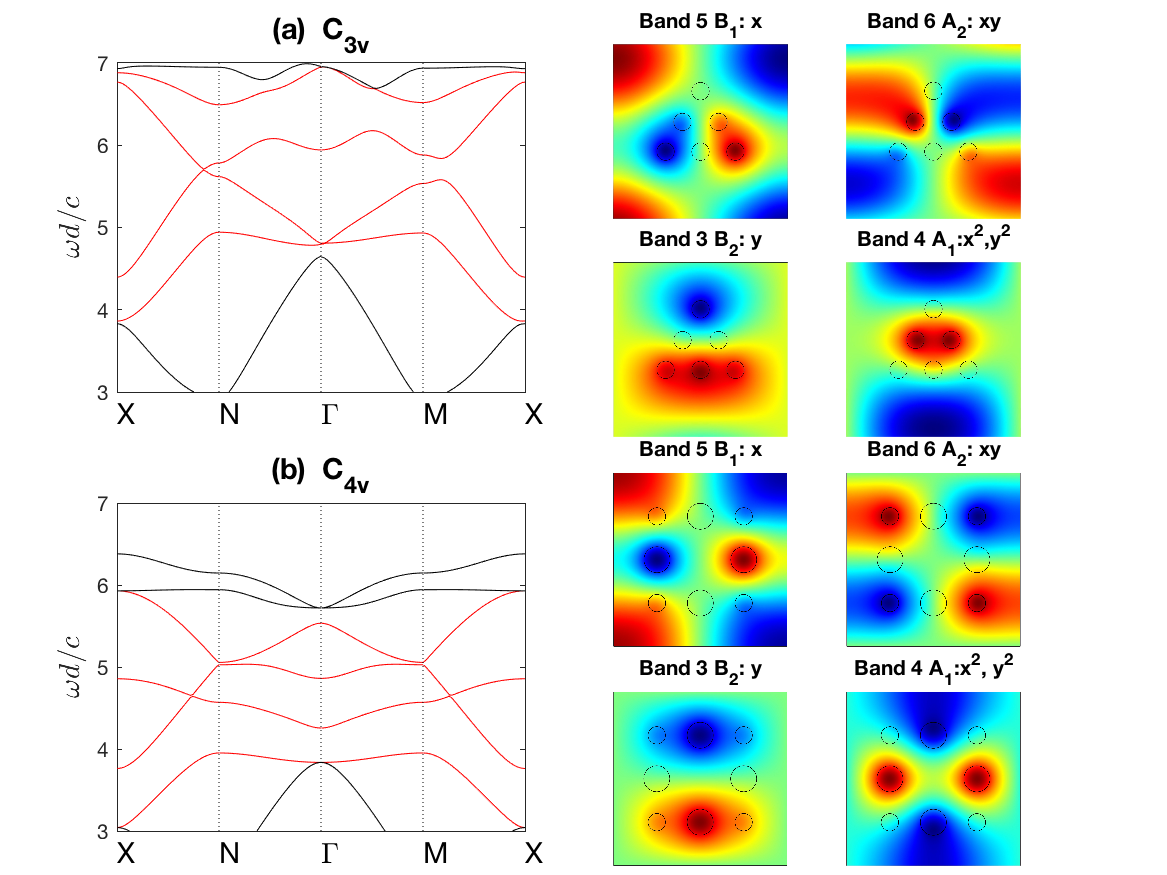}
\caption{
The band diagrams for (a) $C_{3v}$ triangular and (b) $C_{4v}$ square arrangement of inclusions; the red curves show the branches of most interest, bands 3-6, in red; the accidental degeneracy is clear along $XN$ in (a) and along both $XN$ and $MX$ in (b). Also shown are the eigenstates for the bands of interest at $N$. 
}
\label{fig2}
\end{figure}

\begin{figure}[ht!]
\mbox{}
\includegraphics[width=12cm]{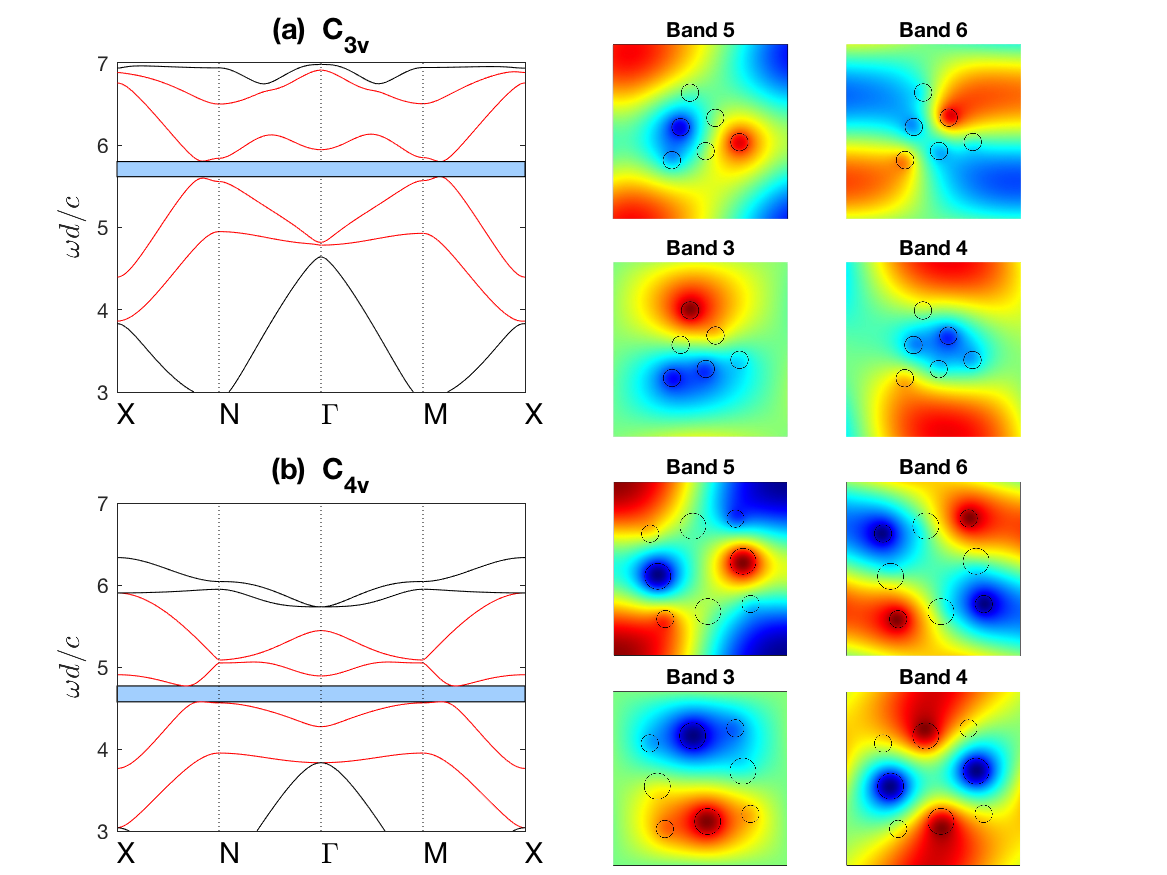}
\caption{
The band diagrams for (a) $C_{3v}$ triangular and (b) $C_{4v}$ square rotated arrangement of inclusions; the red curves show the branches of most interest, bands 3-6, in red. The inclusion sets are rotated anti-clockwise through $\pi/12$ in (a) and $\pi/10$ in (b). 
The associated eigenstates, at $N$, show the orbitals and parity. 
}
\label{fig2_perturbed}
\end{figure}

We begin by considering infinite periodic media and interpret the dispersion diagrams of Fig. \ref{fig2} that contrast the triangular, $C_{3v}$, and square, $C_{4v}$, cases. We take the inclusion arrangements shown in Fig. \ref{fig:Cells_IBZ}(a,b); when we rotate the inclusion arrangements, the rotation is taken around the centroid. 
The most notable difference between the non-rotated cases, Fig. \ref{fig2}(a) and (b), are the single Dirac point for bands 4 and 5 in (a) vis-a-vis two in (b); the most important bands are coloured red and correspond to $n=3,4,5,6$ in the index notation we adopt. As advertised in the introduction a symmetry breaking perturbation, in this case a rotation anti-clockwise, gaps Dirac points to create band-gaps as indicated in Fig. \ref{fig2_perturbed}(a,b). A critical issue, of course, is where the accidental Dirac points arise, or indeed whether they arise at all, \cite{sakoda14a} and this requires an analysis of the band structure. 

A minor point is that, due to the precise position of the inclusions, spaced at $0.25d$, the $C_{4v}$ case has an additional glide and reflectional symmetry and there is a Dirac point exactly at $N$ for bands 5 and 6, but this has no influence on the analysis here. 

To understand how the bands interact we consider the eigenfunctions $\cE_{\bkn}(\bx)$, with $n$ the band index and $\bkappa$ the Bloch wavevector, i.e. the Bloch momentum vector in reciprocal space, in the first Brillouin zone such that $\nabla^2 \cE_{\bkn}(\bx)+a({\bf x})\omega^2_{\bkn}\cE_\bkn(\bx)=0$ with $\omega_\bkn$ the eigenfrequency. 
We consider Bloch waves and the eigenstates $\ket{\cE_\bkn}$ relate to periodic eigenstates $\ket{u_\bkn}$, which form a complete basis, via
\beq
\cE_{n \bkappa}(\bx) = \braket{\bx|\cE_{\bkn}} = \exp\left(i\bkappa\cdot\bx \right) \braket{\bx|u_{\bkn}},
\label{eq:Bloch_condition}
\eeq
with the eigenstates orthonormal, i.e. $
\sum_{n \bkappa} \ket{\cE_{\bkn}} \bra{\cE_{\bkn}} = \hat{1}$,
$\braket{\cE_{\bkn}|\cE_{\bkmprime}} = \delta_{mn} \delta_{\bkappa \bkappa'}$.  

The completeness of the periodic eigenstate basis, means we can expand $\ket{u_{n\bkappa}}$ in terms of the basis set $\{u_{j\bkappa_0}(\bx) \}$ where $\bkappa_0$ is fixed, to deduce
\begin{multline}
\ket{\cE_{n\bkappa}} =  \exp\left(i\bkappa\cdot\bx \right) \ket{u_{n\bkappa}} 
=  \exp\left(i\bkappa\cdot\bx \right) \sum_m A_{nm}(\bkappa) \ket{u_{m\bkappa_0}}  \\ =
\exp\left(i\Delta\bkappa\cdot\bx \right) \sum_m A_{nm}(\bkappa) \ket{\cE_{m\bkappa_0}},
\label{eq:psi_expansion}
\end{multline} 
where $\Delta\bkappa = \bkappa - \bkappa_0$ and $m$ running over all positive integers. Assuming the perturbation 
$|\Delta\bkappa|\ll 1$ the governing equation becomes
\beq
\sum_m A_{nm}(\bkappa)\left[{a(\bx)} (\omega_{n\bkappa}^2 - \omega_{m\bkappa_0}^2) 
+2i \Delta\bkappa \cdot \nabla_{\bx}
+ \mathcal{O} \left(|\Delta\bkappa|^2 \right)\right]\cE_{m\bkappa_0}(\bx)=0.
\label{eq:expansion_gov}\eeq
This relationship connects the eigenstates at a given point in reciprocal space, $\bkappa_0$, to the frequency, and hence determines the local dispersion relation, and can be used to investigate band repulsion or attraction. We concentrate upon the behaviour near the point $N$ as both sets of inclusions share $\sigma_v$ mirror symmetry; the $C_{3v}$ case does not have $\sigma_h$ symmetry and this leads to the "missing" Dirac point along $MX$ in Fig. \ref{fig2}(a). 

Looking further at the band diagrams, say that of Fig. \ref{fig2}(a) one observes, at the wavenumber $N$, close to the Dirac point of interest, that the eigenstates of bands $3-6$ have characteristic orbitals familiar from quantum mechanics  \cite{atkins05a}. The symmetries inherent in these eigenstates motivate the use of the mathematical language of symmetry, that is, group theory:   
The point group symmetry of the structure 
 is $G_{\Gamma} = C_{2v}$; this is also the point group symmetry at $N$, $G_N = C_{2v}$ (Table \ref{tab:C2v}). The $C_{2v}$ point group arises from a combination of spatial (reflectional) and time-reversal symmetries; the latter relates $\bkappa \rightarrow -\bkappa$.
  The irreducible representations (IRs) at  $N$ are one-dimensional hence there is no symmetry induced degeneracy; however, two of the IRs can be tuned such that an accidental degeneracy, that is not symmetry repelled, is created along $NX$ (and also along $MX$ for the $C_{4v}$ case). 

The four bands highlighted in red in Fig. \ref{fig2}(a,b) (bands numbered $3-6$ inclusive) are associated with the eigenstates shown alongside the dispersion curves; these match the basis function symmetries of the $C_{2v}$ group given in Table \ref{tab:C2v}(a); hence this indicates that bands $3-6$ are symmetry induced and the sequential ordering of them (lowest to highest) is deduced numerically, via the eigenstates, as: $\{B_2, A_1, B_1, A_2\}$ and these are shown in Fig. \ref{fig2}. The rotated, symmetry-breaking, cases of Fig. \ref{fig2_perturbed}(a,b) retain, at least broadly, the same eigenstate structure. 

\begin{table}[h!]
    \begin{minipage}{.1\linewidth}
      \centering
       \caption*{(a)}
\begin{tabular}{|l|cccc|c|}
\hline
\cellcolor[HTML]{EFEFEF}Classes $\rightarrow$ &                       &                         &                              &                              &                                                                              \\
\cellcolor[HTML]{EFEFEF}IR $\downarrow$       & \multirow{-2}{*}{$E$} & \multirow{-2}{*}{$C_2$} & \multirow{-2}{*}{$\sigma_v$} & \multirow{-2}{*}{$\sigma_h$} & \multirow{-2}{*}{\begin{tabular}[c]{@{}c@{}}Basis \\ functions\end{tabular}} \\ \hline
$A_1$                                         & $+1$                  & $+1$                    & $+1$                         & $+1$                         & $x^2, y^2$                                                                   \\
$A_2$                                         & $+1$                  & $+1$                    & $-1$                         & $-1$                         & $xy$                                                                         \\
$B_1$                                         & $+1$                  & $-1$                    & $+1$                         & $-1$                         & $x, xy^2$                                                                    \\
$B_2$                                         & $+1$                  & $-1$                    & $-1$                         & $+1$                         & $y, x^2y$                                                                    \\ \hline
\end{tabular}
\end{minipage} 
\begin{minipage}{1.35\linewidth}
      \centering
       \caption*{(b)}
\begin{tabular}{|l|cc|c|}
\hline
\cellcolor[HTML]{EFEFEF}Classes $\rightarrow$ &                       &                                   &                                                                              \\
\cellcolor[HTML]{EFEFEF}IR $\downarrow$       & \multirow{-2}{*}{$E$} & \multirow{-2}{*}{$\sigma_{v, h}$} & \multirow{-2}{*}{\begin{tabular}[c]{@{}c@{}}Basis \\ functions\end{tabular}} \\ \hline
$A$                                           & $+1$                  & $+1$                              & $x^2, y^2, xy$                                                               \\
$B$                                           & $+1$                  & $-1$                              & $x, y, x^2y, xy^2$                                                           \\ \hline
\end{tabular}
    \end{minipage} 

\caption{Character table showing the classes, irreducible representations (IRs) and basis functions for (a) $C_{2v}$ and (b) reflections $\sigma_{v,h}$.}
\label{tab:C2v}
\end{table}

Intuitively, 
 we expect the two bands (band 4 and 5)  forming the accidental degeneracy, $A_1, B_1$, to be strongly coupled with each other; the other two symmetry induced bands, $B_2, A_2$, will have limited influence on the local curvature or slope of the $A_1, B_1$ bands. The even more distant bands (bands 1 and 2 and those above 6) will have negligible effect on  the $A_1, B_1$ bands and we quantify this by separating the bands into the 
 the symmetry set eigenstates (SSE, bands $3-6$) and the remainder that lie outside the SSE. 
 Eq. \eqref{eq:psi_expansion} becomes 
\beq
\ket{\cE_{n\bkappa}} =  
\exp\left(i\Delta\bkappa\cdot\bx \right) \bigg[ \sum_{m \in \text{SSE}} A_{nm}(\bkappa) \ket{\cE_{m\bkappa_0}} + \sum_{m \notin \text{SSE}} A_{nm}(\bkappa) \ket{\cE_{m\bkappa_0}}  \bigg]. 
\label{eq:psi_separation} 
\eeq 
Multiplying \eqref{eq:expansion_gov}, by the conjugated states $\cE_{p \bkappa_0}^* (\bx)$ or $\cE_{q \bkappa_0}^* (\bx)$ (where $p \in \text{SSE}, q \notin \text{SSE}$) and integrating over the primitive cell in physical space  we obtain the following two equations,
\beq
\begin{split}
(\omega_{n\bkappa}^2-\omega_{p\bkappa_0}^2)A_{np} &=  \sum_{m \in \text{SSE}}H_{pm}A_{nm} + \sum_{m\notin\text{SSE}}H_{pm}A_{nm}, \\
(\omega_{n\bkappa}^2-\omega_{q\bkappa_0}^2)A_{nq} &=  \sum_{m\in\text{SSE}}H_{qm}A_{nm} + \sum_{m\notin\text{SSE}}H_{qm}A_{nm}
\label{eq:simul_eqns}
\end{split}
\eeq 
and the $H_{ab}$ are explicitly,
\beq
H_{ab} = -2i \Delta\bkappa \cdot \braket{\cE_{a\bkappa_0}|
\frac{\nabla_{\bf x }}{a({\bf x})}
|\cE_{b\bkappa_0}} + \mathcal{O}\left(|\Delta\bkappa|^2 \right)
\label{eq:aug_Ham} 
\eeq 
where we recall that here the function $a({\bf x})$ is piecewise constant. 
A useful point is that, using the symmetries of the eigenstates one can select which $H_{ab}$ are zero. 
Neglecting the terms coupling states outside the SSE to each other we obtain, for $p\in$ SSE, that 
\beq
(\omega_{n\bkappa}^2-\omega_{p\bkappa_0}^2)A_{np} =  \sum_{m\in\text{SSE}} A_{nm}\left(H_{pm} +
 \sum_{q\notin\text{SSE}} \frac{H_{pq} H_{qm}}{(\omega_{n\bkappa}^2-\omega_{q\bkappa_0}^2)}\right).
\label{eq:total_equation}
\eeq
Our main interest is in the neighbourhood of the Dirac point so we 
 set $n=p \in \text{SSE}$, and perturb with $\bkappa = \bkappa_0 + \Delta\bkappa$, then 
\beq
\omega_{n\bkappa}^2 = \omega_{p\bkappa_0}^2 + 2\omega_{p\bkappa_0} \Delta\bkappa \cdot \nabla_{\bkappa} \omega_{p\bkappa_0} + \mathcal{O}\left(|\Delta\bkappa|^2\right).
\label{eq:freq_exp}\eeq
Therefore the second summation in \eqref{eq:total_equation}, coupling states within the SSE to those outside, falls into second-order, i.e. $\mathcal{O}\left(|\Delta\bkappa|^2\right)$, hence the first-order equation is the $4\times 4$ system, 
\beq
 \left(2\omega_{p\bkappa_0} \Delta \omega_p \right)A_{np} = \sum_{m \in \text{SSE}}H_{pm}A_{nm}. 
\label{eq:effective_eqn}\eeq 
where $ \Delta \omega_p =\omega_{p\bkappa} - \omega_{p\bkappa_0}$ and $n,p \in \text{SSE}$.
 Equation (\ref{eq:effective_eqn}) contains the information that allows us to determine whether an accidental Dirac degeneracy will occur. One can take this further and notably, the higher-order corrections, that encompass the coupling between bands within the SSE to those outside, provide the band curvature details away from a locally linear point. In that instance, Eq. (\ref{eq:effective_eqn}) is modified to a $4\times4$ matrix eigenvalue problem, where the Hamiltonian, with components $H_{pm}$, is Hermitian \cite{he15a}.


\subsection{Compatibility relations and creating accidental degeneracy along $NX$}

Bands vary continuously, except possibly at accidental degeneracies where mode inversion may occur, which in turn leads to a discontinuity of the intersecting surfaces. Hence when moving along a continuous band of simple eigenvalues the eigenstates continuously transform; departing from the high symmetry point $N$, the associated IRs describing the transformation properties of the eigenstates smoothly transition into IRs that belong to the point groups along $N\Gamma$ or $NX$.  

 In physical space both cellular structures we consider in Fig. \ref{fig:Cells_IBZ} have $\sigma_v$ spatial symmetry, this is equivalent to $\sigma_h$ symmetry in Fourier space. From the definition of a point group symmetry, i.e. any symmetry operator $\hat{R} \in G_{\Gamma}$ that satisfies, $\hat{R} \bkappa = \bkappa \mod {\bf G}$, where ${\bf G}$ is a reciprocal lattice basis vector, implies that $\bkappa \in NX$ solely has the mirror symmetry operator, $\sigma_h$ within its point group. Similarly, for a $\bkappa \in \Gamma N$, only the vertical mirror symmetry operator, $\sigma_v$
  satisfies the point group criterion. The symmetries of the eigenstates, for a $\bkappa$ belonging to either of the paths, $NX$ and $N\Gamma$, are shown within the basis functions column of Table \ref{tab:C2v}(b). 
 
 We now simplify the $4\times 4$ system of Eq. \eqref{eq:effective_eqn} by assuming it is dominated by the two strongly coupled bands, bands 4 and 5 with IRs $A_1$ and $B_1$, which will result in a $2\times 2$ system.   
 As we move away from $N$, the $A_1, B_1$ IRs belonging to the $C_{2v}$ table, from continuity of the bands, transform into the IRs of the $\sigma_{v,h}$ table and reveal compatibility relations of the IRs. For the  symmetry $\sigma_h$ the eigenstates at $N$ and along $NX$ satisfy the following,
\beq
\hat{P}_{\sigma_h} \ket{\cE_{A_1, B_1}} = \pm  \ket{\cE_{A_1, B_1}}, \quad \hat{P}_{\sigma_h} \ket{\cE_{A, B}} = \pm  \ket{\cE_{A, B}}
\label{eq:sigma_h} \eeq
 where $\hat P$ is the projection operator. 
The bands $(A_1, B_1)$ at $N$ are therefore compatible with $(A, B)$ along $NX$. Physically, this transition is also evident from the eigenstates. 
Similarly, at $N$ and along $N\Gamma$, the eigenstates transform under $\sigma_v$ as,
\beq
\hat{P}_{\sigma_v} \ket{\cE_{A_1, B_1}} = +  \ket{\cE_{A_1, B_1}}, \quad \hat{P}_{\sigma_v} \ket{\cE_{A, B}} = \pm \ket{\cE_{A, B}}.
\label{eq:sigma_v}
\eeq
The bands $(A_1, B_1)$ at $N$ are therefore compatible with $(A, A)$ along $N\Gamma$ which implies band repulsion and an inability to have a band crossing along $N\Gamma$ as we observe. 

 Importantly, note that, in deriving equation \eqref{eq:effective_eqn} we have only assumed that $\bkappa_0$ belongs to a particular symmetry set band (surfaces $3-6$) (the band at $\bkappa_0$ must be continuously connected to the same band at $N$). Therefore, the compatibility relations allow us to choose any expansion point along the the path $\Gamma N X$ where the eigenfunction basis set, \eqref{eq:psi_expansion}, transforms accordingly i.e. $\ket{\cE_{A_1}} \rightarrow \ket{\cE_{A}}$.

\begin{figure}[!h]
    \centering
    \captionsetup{justification=raggedright}
    \begin{minipage}{.245\textwidth}
    \hspace{-2.75cm} 
    \caption*{(a)}
        \includegraphics[width=0.55\linewidth]{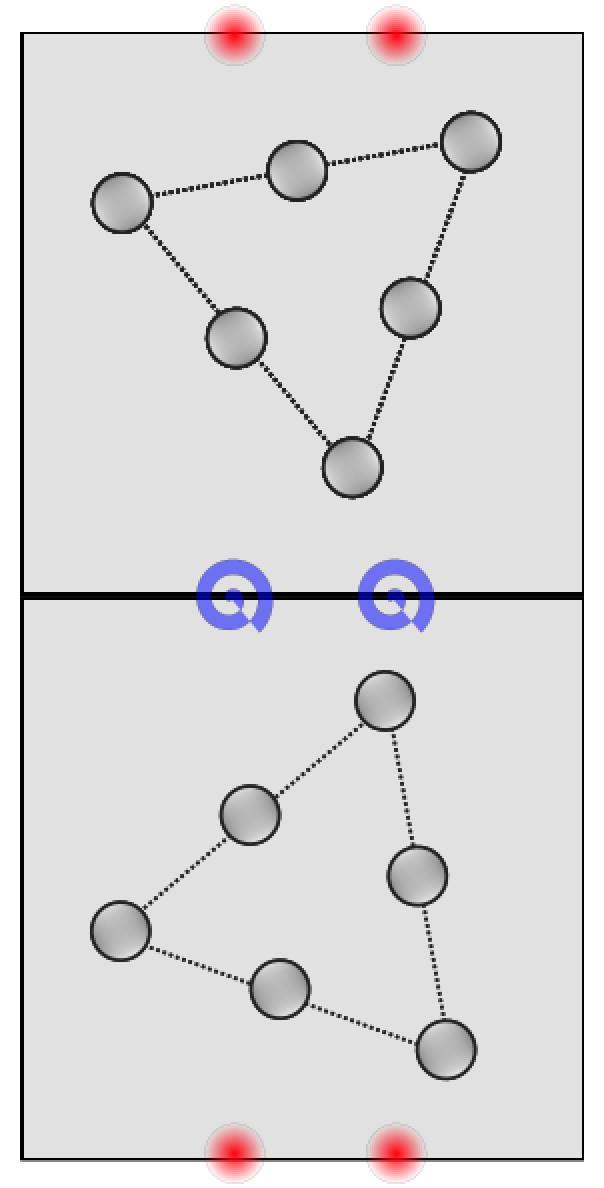}
    \end{minipage}%
    \begin{minipage}{0.245\textwidth}
        \hspace{-1.75cm}
        \caption*{(b)}
           \hspace{-0.75cm}
        \includegraphics[width=0.55\linewidth]{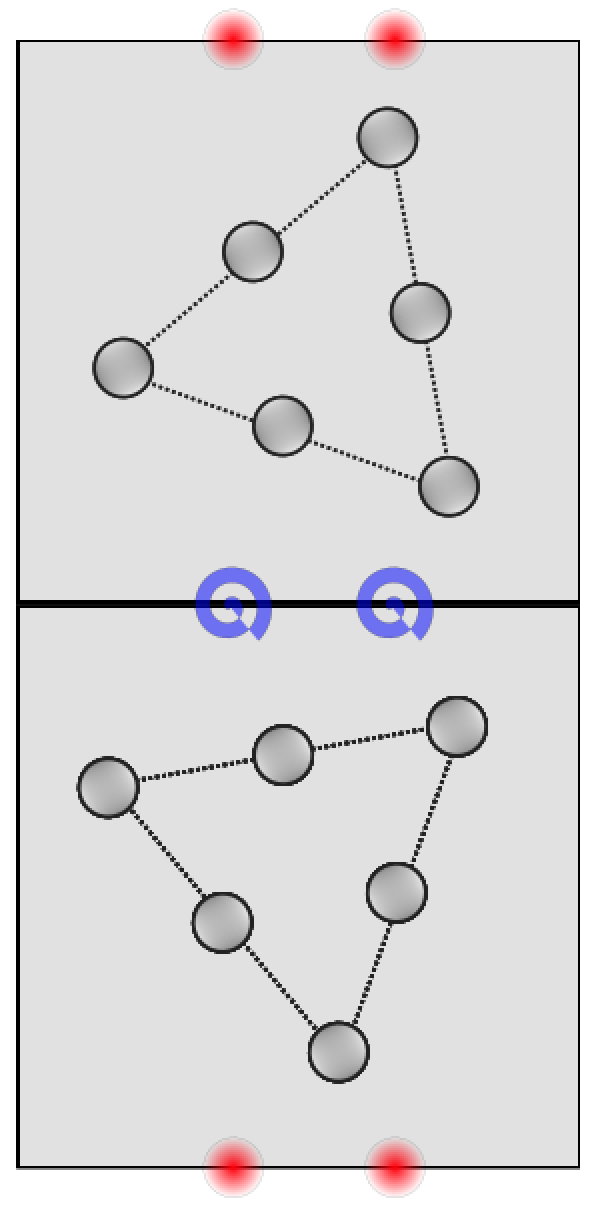}
    \end{minipage}
     \begin{minipage}{0.245\textwidth}
      \caption*{(c)}
        \hspace{-1.45cm}
                \includegraphics[width=1.15\linewidth]{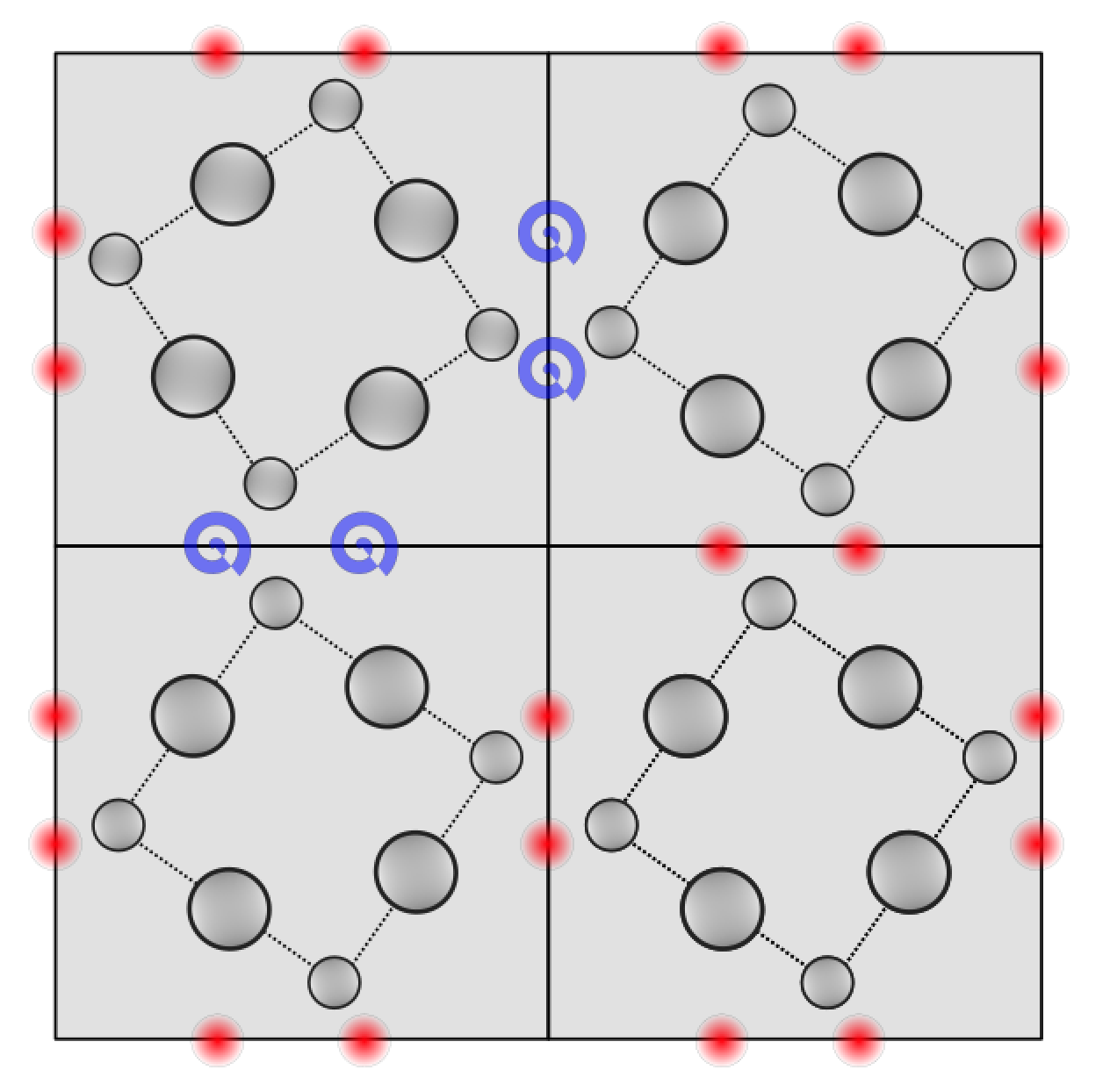}
    \end{minipage}
        \begin{minipage}{0.245\textwidth}
        \caption*{(d)}
        \hspace{-0.65cm}
        \includegraphics[width=1.15\linewidth]{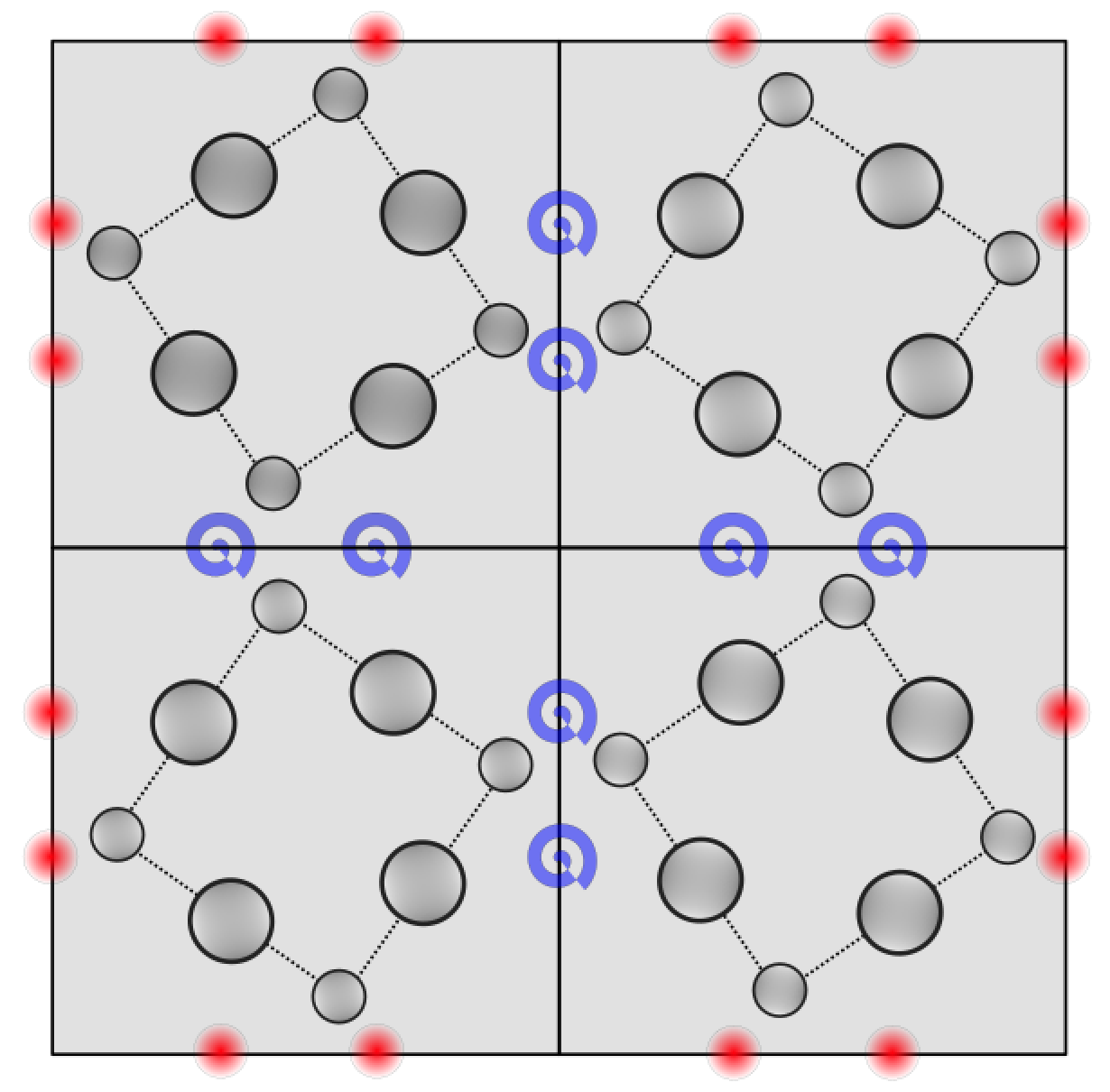}
    \end{minipage}
    \caption{The interfaces created when regions of perturbed media neighbour its counter-rotated twin. 
    Panels (a) and (b) show, for the triangular $C_{3v}$ case, that there are two distinct interfaces where the triangles in opposing media are either point-to-point or face-to-face. 
    The square arrangement of inclusions, $C_{4v}$, is shown in (c) where the top left medium is oppositely orientated to the others (used for propagation around the simple bend) and (d) where the top left and bottom right media are oppositely orientated to the other quadrants (used for propagation in a splitter). Red dots show interfaces with nonzero Berry curvatures, whilst purple spirals show an overlap between regions with opposite Berry curvature where ZLMs can reside.}
    \label{fig:Cell_Arrangement}
\end{figure}

\begin{figure}[ht!]
\mbox{}
\centering\includegraphics[width=14cm]{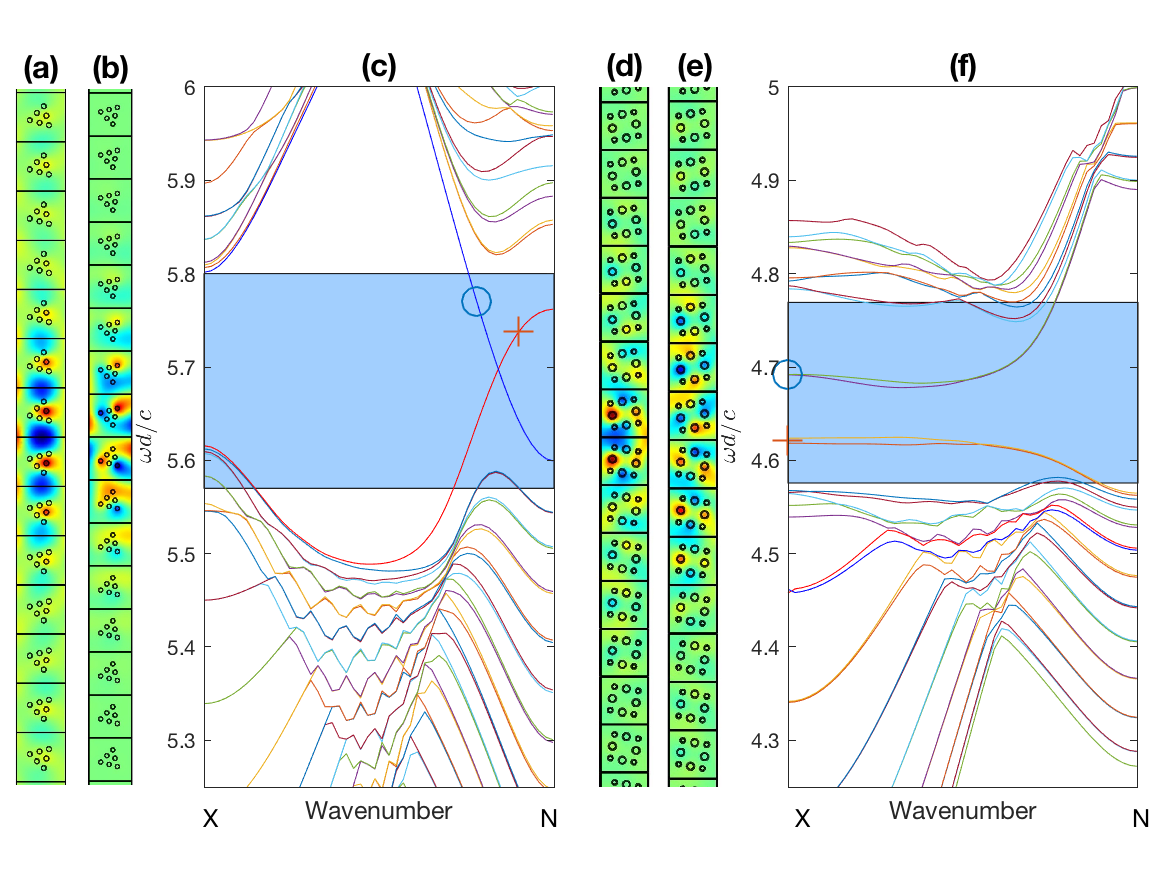}
\caption{Edge states (ZLMs):  For the interface between oppositely tilted triangles (a-c), and squares (d-f). The even and odd  modes are shown in (a,d) and (b,e) respectively at normalised frequencies given by the circles, $(5.77, 4.69)$,
and crosses, $(5.74,4.62)$,
in the dispersion curves of (c,f).
}
\label{fig:strip}
\end{figure}

In order to solve the 2-band eigenvalue problem, Eq. \eqref{eq:effective_eqn}, we compute the determinant of the truncated matrix,
\beq
\begin{vmatrix}
    \omega_{A_1} \Delta \omega_{A_1}     &  -\Delta\kappa_x \braket{\psi_{A_1}|i\partial_x/a|\cE_{B_1}} \\
   - \Delta\kappa_x \braket{\psi_{A_1}| i\partial_x/a|\cE_{B_1}}^*      & \omega_{B_1} \Delta \omega_{B_1}  
\end{vmatrix} = 0, 
\eeq 
where parity simplifies the Hermitian matrix; the eigenstates are evaluated at $\bkappa_0$.
 Solving the eigenvalue problem yields the following result,
\beq
\omega_{A_1, B_1} \Delta \omega_{A_1, B_1} = \pm    \vert\Delta\kappa_x \braket{\cE_{A_1}|i \partial_x/a|\cE_{B_1}}\vert,
\label{eq:local_linear}\eeq
where the $\pm$ corresponds to the $A_1, B_1$ bands, respectively.
This result implies that the $A, B$ bands have an identical slope, albeit with opposite gradients; hence, if, at $N$ an instance can be found where $\omega_{B_1} > \omega_{A_1}$  then the bands will invariably cross along the path $NX$. We are not guaranteed that an accidental degeneracy must occur along $NX$ as parameters (the radii of inclusions, number of inclusions, permittivity etc) could occur with $\omega_{B_1}<\omega_{A_1}$, but this inequality gives a useful criteria for their existence, or otherwise. This parametric freedom afforded by  inclusion change in geometry or material  
 tunes increases, or decreases, in the slope thereby increasing or decreasing the distance between $N$ and the Dirac point.
 
 Note that the Dirac cone occurs along the spatial symmetry path, $\sigma_h$, of the structure due to the opposite parities of the $A, B$ bands; band repulsion occurs along the $N \Gamma$ path \cite{he15a} thereby resulting in a partial band gap along $N\Gamma$. If $\omega_{B_1} > \omega_{A_1}$, then the partial gap along $N\Gamma$ isolates the Dirac cone along a portion of the IBZ path, $\Gamma N X$.

We have designed situations where either two, or four, (for $C_{3v}$ and $C_{4v}$ respectively) pairs of Dirac points are created by an accidental degeneracy and we have shown how they are created and given a prescription for their occurence. These Dirac points have been gapped by a symmetry breaking perturbation and band gaps have been created.

\section{Results and discussion}

\subsection{Edge states: Zero line modes}

\begin{figure}[ht!]
\mbox{}
\centering\includegraphics[width=13cm]{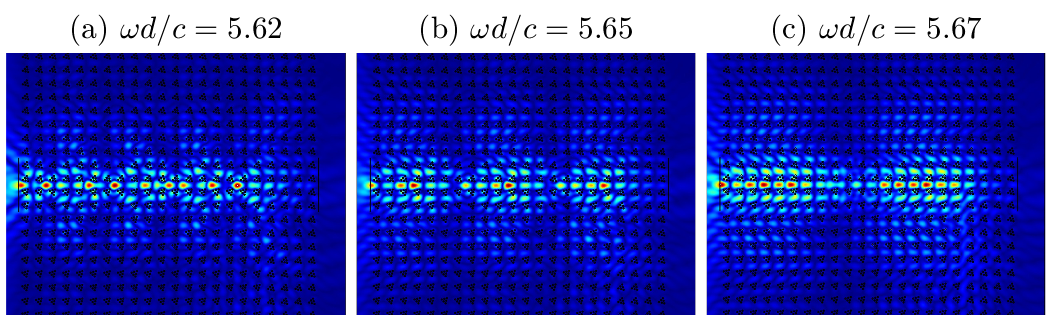}
\caption{ZLMs along the face-to-face interface between media composed of oppositely tilted triangles at normalised frequencies $5.62, 5.65, 5.67$ illustrating the long-scale envelope and concentration of field to the interface. 
}
\label{fig:ZLM}
\end{figure}

\begin{figure}[ht!]
\mbox{}
\centering\includegraphics[width=13cm]{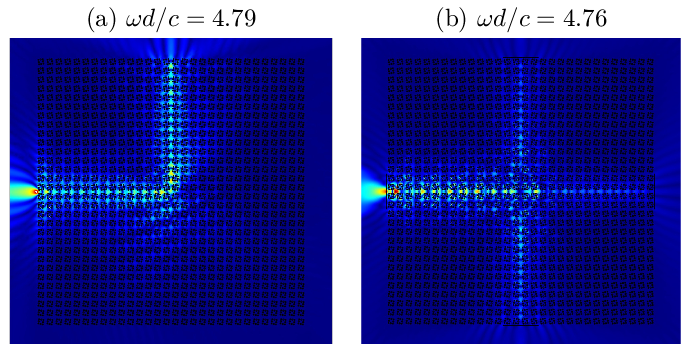}
\caption{(a) Propagation around a $\pi/2$ bend at the interface between a quadrant of squares oppositely orientated to the others, c.f. Fig. \ref{fig:Cell_Arrangement}(c). (b) The three-way splitter, where the quadrants alternate in their relative rotation, c.f. Fig. \ref{fig:Cell_Arrangement}(d). In both panels the excitation is a monopole source located at the leftmost edge of the crystal on the interface between the media.
}
\label{fig:bend}
\end{figure}

We aim to use our knowledge of the designed band-gaps to create situations whereby an interface will support edge states. We begin by taking a half-space of one medium and place it above another; the only difference between the two media being that the symmetry of the cells in the upper and lower media is broken by clockwise and counter-clockwise rotations respectively (see Fig. \ref{fig:Cell_Arrangement}(a,b) for $C_{3v}$). This has the effect, in Fourier space, of interchanging $N$ and $M$, the Berry curvatures \cite{xiao10a} are opposite in sign at $N$ and $M$, and hence at the interface between such media and create valley Hall edge states; these are named zero line modes (ZLMs) due to their origin as arising from opposite Berry curvature in the adjoining media. 

We compute ZLMs by considering finite tall ribbons one cell thick with Bloch conditions applied on the sides and with periodic conditions at the very top and bottom; the ribbons in the simulations are $40$ cells in length which is sufficient to ensure that the exponential decay of the edge states does not interact with the precise boundary conditions at the top and bottom of the ribbon. We extract decaying edge states, even and odd, around the interface where, recalling the polarisation chosen, the even modes are the physically relevant ones. The eigenstates and dispersion curves are shown in in Fig. \ref{fig:strip} and the parity of the ribbon eigenstates is inherited from the bulk eigenstates of Fig. \ref{fig2}, \ref{fig2_perturbed}. 
Notably the $C_{3v}$ and $C_{4v}$ cases have a crucial difference: ordering of the media matters, that is, stacking the clockwise above the counter-clockwise or vice-versa leads to two different interface types for the triangular case whilst for the square case it is irrelevant which ordering is taken (see Fig. \ref{fig:Cell_Arrangement}). The differences in the interfaces in the $C_{3v}$ versus $C_{4v}$ cases form a key distinction that impacts upon energy transport around corners and splitting. 

Full scattering simulations performed using the commercial finite element package COMSOL \cite{comsol} are shown in  Fig. \ref{fig:ZLM}, these are for the even ZLM relevant for TM polarisation and for the $C_{3v}$ case; very similar ZLMs are found for $C_{4v}$ (not shown). The excitation is a line source just outside the leftmost edge of interface and a very clear ZLM is excited that can be identified, on each ribbon, with the eigenstate from the ribbon. One feature that is also evident is the long-scale wave envelope with a wavelength that alters with frequency; this can be described via an effective medium approach \cite{antonakakis14a} as applied to edge states \cite{makwana19a}. 

\subsection{Energy transport around a sharp bend and energy splitting}

We now consider two related, yet distinct, problems: redirecting energy around a sharp bend using just topology, see Fig. \ref{fig:bend}(a), which is the topological alternative to the photonic crystal waveguides pioneered in \cite{mekis96a}, and a three-way energy splitter (see Fig. \ref{fig:bend}(b)). Both panels in Fig. \ref{fig:bend} are for inclusions placed around a square, the $C_{4v}$ case; the $C_{3v}$ case, despite creating a clear ZLM as in Fig. \ref{fig:ZLM}, is incapable of supporting a ZLM along the vertical interface as those interfaces, c.f. Fig. \ref{fig:Cell_Arrangement}, do not have non-zero Berry curvature. The square arrangement of inclusions has the very useful property that the vertical interfaces are exactly the same as the horizontal ones; the added benefit of having two reflectional symmetries in the $C_{4v}$ case is now evident there is now  non-zero Berry curvature along both interfaces and both support the same ZLMs. It is this insight that allows for the design of the splitter and allows transport around the bend.  

We now want to optimise the transport properties. First, we may wish to minimise backscatter from the junction. To do so we note that the Fourier separation between the Dirac point of the unperturbed bulk dispersion curves and high-symmetry point is highly relevant for the transmission properties of the topological guide \cite{makwana18b}; transmission improves for short wavelength, as opposed to long wavelength, envelopes, hence, for transmission post the junction, it is desirable to increase the distance between the Dirac cone and the point $N$. The latter holds due to the connection between the bulk and projected bandstructures \cite{bostan_design_2005}; the Brillouin zone reduces to one-dimension because the only relevant wavevector component is along the interface. All wavevectors are projected onto the $\Gamma N$ line in Fourier space, hence if the distance between $N$ and the Dirac cone is increased then the Fourier separation between oppositely propagating modes, along the topological guide, is increased.
A mechanism to do this is to alter the system parameters; Eq. \eqref{eq:local_linear} demonstrates that the slopes of the $A$ and $B$ bands can be increased or decreased by the system parameters thereby altering the position of the band intersection.

\begin{figure}[ht!]
\mbox{}
\centering\includegraphics[width=8cm]{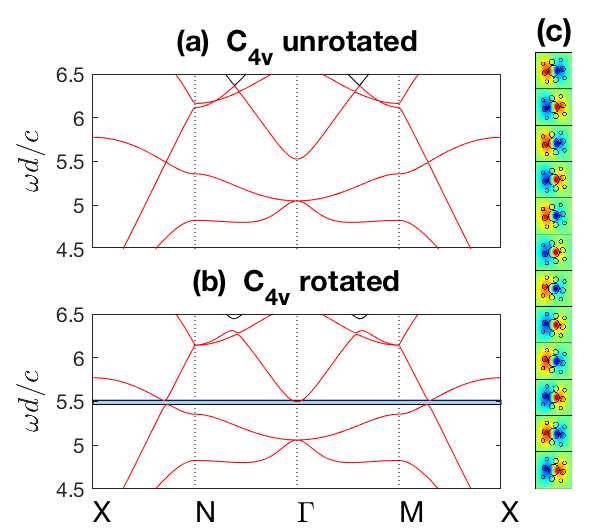}
\caption{The magnetic, TE, case. (a) shows a $C_{4v}$ case, (b) rotated by $\pi/10$, and (c) the odd ZLM eigenstate (at normalised frequency $5.43$). 
}
\label{fig:magnetic}
\end{figure}

Second, we return to envelope wavelength and note that the distance from the source to the junction will play a role. If we took a finite length slab then fitting an integer, or half integer, numbers of envelope wavelengths along the lead interface gives a Fabry-Perot resonance with perfect transmission or perfect reflection from the far edge. We can use this knowledge to tune the system and the sharp bend is optimised by having a node of the long-scale envelope at the junction and so the energy is smoothly transported around the corner. Whilst for the splitter, the perfect reflection scenario concentrates energy at the junction for subsequent redistribution to the exit leads. 

This is the first example of a three-way splitter passively created due to its topology and as a result of the inherited protection should be less prone to backscatter and therefore forms the prime candidate for three-way splitting in time reversal systems. 


\section{Concluding remarks}

We have constructed a three-way splitter, and designed for propagation around a right angle bend, using valleytronics and the presence of accidental Dirac points. Although we have concentrated upon the TM polarisation, it is clear that TE polarisation will also generate splitters using the geometrical designs here; the main change being that the odd (and not even) modes, now excited by dipoles, would be the physically relevant fields. We briefly illustrate this in Fig. \ref{fig:magnetic} for a square $C_{4v}$ case; here the inclusions of Fig. \ref{fig:Cells_IBZ}(b) are augmented by a central inclusion of radius $0.15d$. The situation is almost identical to TM, except that the bandgap is smaller (the action of the central inclusion is to help create the bandgap) and the decay of the edgestate, shown in Fig. \ref{fig:magnetic}(c), is slower. 

One crucial difference from the majority of the valleytronics literature is that we have chosen to operate on a square, and not hexagonal, lattice; the hexagon arrangement has several advantages - the Dirac points are symmetry induced, and the band-gaps obtained by gapping them can be constructed to be broad. For energy transport from one interface to another both the chirality and phase must match; this is easily achieved for the square arrangement designed here as the interfaces match. For four joined quadrants of structured hexagonal lattice there is a mismatch in phase and so you are confined to two-way splitting and thus the square splitter here appears the only option.

\section*{Acknowledgments}
The authors thank the EPSRC for their support through grant
{EP/L024926/1} and R.V.C acknowledges the support of a Leverhulme Trust
Research Fellowship and US Air Force Office of Scientific Research / EOARD (FA9550-17-1-0300).

\end{document}